# SUCCESS OF CHEMOTHERAPY IN SOFT MATTER

I. Trifonova[1], G. Kurteva[1], S. Z. Stefanov[2]

[1]*National Oncology Medical Center, 6 Plovdivsko Pole Str., Sofia 1756, Bulgaria*

[2]*ESO EAD, 5 Veslets Str., 1040 Sofia, Bulgaria*

E-mails: itrifa@abv.bg; dr.kurteva@gmail.com; szstefanov@ndc.bg

**Abstact.** The success of chemotherapy in soft matter as a survival is found in the paper. Therefore, it is found the analogous tumor stretching force in soft matter; ultrasonography is performed for this tumor; restoration in soft matter with such a tumor is found; Bayes estimate of the probability of chemotherapy success is derived from the transferred chemical energy and from soft matter entropy; survival probability is juxtaposed to this probability of success.

## 1. Introduction

Soft matter is a matter between the nano and micron scales, where the new characteristics occur, due to the collective behavior in it and as well the complex flow there [1]. The scale of this matter is the same as the scale of life [2].

Cancer development is a tumor growth [3]. When human tissues are considered as soft matter, cancer development is not only a growth but also an intensive thermodynamic process. This intensive process causes molecule knotting, changing of molecule elasticity and new molecule networks. Cancer is a dynamic phase transition [4] in such an intense process.

The proliferation index is a clinical prognostic biomarker of cancer development. It estimates the expected doubling time of tumors. It is obtained by investigating PCNA protein (proliferating cell nuclear antigen).

The survival [5-6] demonstrates the success of cancer chemotherapy. Here, the success of chemotherapy is considered as a ratio of the efficacy to it toxicity.

In this paper is studied the success of chemotherapy in soft matter through the transferred chemical energy in soft matter. Here, the transferred chemical energy is obtained from the proliferation index.

## 2. Analogous force of tumor stretching

Tumor proliferation in soft matter determines the reaction temperature between the tumor and soft matter. The inverse temperature of the reaction β is obtained according to Tian [7] from the proliferation index

$$\beta = \pi(PI_1^2 - 0.36)/(10PI) \qquad (1)$$

In (1) PI is the proliferation index in percentages, divided by 10.

Cancer proliferation periodicity in soft matter is determined by the inverse temperature β in accordance with Cardy [8]

$$\omega = 4/\beta, \qquad (2)$$

In (2) ω is the cancer proliferation frequency.

Let the tumor in soft matter acts as stretching in an isothermal system from [9]. Then the analogous stretching force F is found from cancer proliferation frequency ω,

$$F = -0.4343\ln(0.4343)/(1+1/\omega) \qquad (3)$$

## 3. Flow in Tumor Development

The analogous stretching force changes the internal soft matter energy.

Initial internal energy of soft matter is obtained according to [10] as an initial internal energy of a quantum thermodynamic system with a preliminary information about the tumor stretching force. This initial internal energy $E_{ini}$ is

$$E_{ini} = ((a_{max} - a_{min}) + T_2\ln((1 + \exp(a_{min}/T_2))/ (1 + \exp(a_{max}/T_2))))/\ln(a_{max}/a_{min}), \qquad (4)$$
$$T_2 = \min(1/\beta, 1/\beta_0), a_{min} = \min(E, E_0), a_{max} = \max(E, E_0)$$

In this equation β is the inverse tumor temperature, E is the tumor energy, $\beta_0$ is the base inverse tumor temperature, $E_0$ is the base tumor energy, and $T_2$ is the temperature of soft matter without a tumor.

The final internal energy of soft matter is obtained according to [10] as a final internal energy of the quantum thermodynamic system with preliminary information for the tumor stretching force. This final internal energy $E_{fin}$ is

$$E_{fin} = 176((a_{max} - a_{min}) + T_2\ln((1 + \exp(a_{min}/T_2))/ (1 + \exp(a_{max}/T_2))))(1 - T_2/T_1) \qquad (5)$$
$$T_1 = \max(1/\beta, 1/\beta_0)$$

Herein β, E, $\beta_0$, $E_0$, $T_2$ are the values from (5), and $T_1$ is the temperature of the soft matter with a tumor. The change of the internal energy of the soft matter at analogous stretching force is

$$S = E_{fin} - E_{ini} \qquad (6)$$

This change in internal energy is heard in ultrasonography [11]. Indeed [11], when ultrasound beam passes through the tissue, its energy is partly absorbed and converted into heat. The transfer of ultrasound energy is characterized by parameters (a, λ, μ), defined according to [11] from the change of the internal energy and the tumor stretching force for L experiments. These parameters are

$$a = \eta B_3/(F(1 - B_1 - B_2)), \lambda = (2 - B_1)/(a + 1 - \eta), \mu = (1 - B_1 - B_2)/(\lambda\eta), \qquad (7)$$
$$\eta = 10(1 + (1 - T_2/T_1)/\ln(T_2/T_1)),$$
$$SB = E,$$

$$S = \begin{vmatrix} S_{11} & S_{12} & S_{13} \\ S_{21} & S_{22} & S_{23} \\ S_{31} & S_{32} & S_{33} \end{vmatrix}, B = \begin{vmatrix} B_1 \\ B_2 \\ B_3 \end{vmatrix}, E = \begin{vmatrix} E_1 \\ E_2 \\ E_3 \end{vmatrix},$$

$S_{ij} = \sum_{k=2..L-2} S_{k-i}S_{k-j}, E_i = \sum_{k=2..L-2} S_k S_{k-i}, i = 1..3, j = 1..3,$
$S_{i-1} = S_i, i = 1..4; S_{i-2} = S_i, i = 6..L$

Herein F is the tumor analogous stretching force from (3), and $S_i$ is the change of the internal energy for the $i^{th}$ experiment, i = 1..L.

Ultrasonography gives the following metric tumor marker $m_n$,

$m_n = \{7.3, \text{ if } N\_int(u_n) \leq 0; 4.1, \text{ if } N\_int(u_n) = 1; 2.9, \text{ if } N\_int(u_n) = 2;$
$2.2, \text{ if } N\_int(u_n) = 3; \text{ undetermined, if } N\_int(u_n) \geq 4 \}$ (8)

$u_n = 100(2.5\mu - a)$

Herein $N\_int(.)$ is the nearest integer to $(.)$.

## 4. Transferred Chemical Energy during Chemotherapy

Resource restoration of soft matter with a tumor is determined by the analogous force of stretching F and by the chemotherapy intensity $k_s$. Then [12], resource restoration of soft matter with a tumor proceeds with the following intensity

$m_a = (2.2/k_s)\exp(2(F+1.05))-3.5$ (9)

Resource restoration of the soft matter with a tumor with intensity (10) is achieved in accordance with [13], with a percent drug W,

$W = 100-50(100k_s/m_a)^{1/2}$ (10)

The time $t_r$ for resource restoration with a percent drug W is obtained according to [13]

$t_r = (42.5+(42.5^2 - 16.16(W+10.7))^{1/2})$ (11)

This time is obtained at isothermality of soft matter with a tumor in it. The change of this time, due to the heat diffusion in the soft matter with a tumor, is [14], [12],

$\Delta t = \sum_{j=1,...,3} (1/g_j - g_j^2/(2g_1g_2g_3))$, (12)

$g_1 = (1/4)(100k_s-m_a)$, $g_2 = (1/4)\ln(100k_s/m_a)$, $g_3 = (1/4)(1/m_a - 1/(100k_s))$

The time $T_r$ for resource restoration of the soft matter with a tumor, at taking into account the heat diffusion, is

$T_r = t_r + \Delta t$ (13)

Transferred chemical energy into the soft matter at chemotherapy with a restoration time (13), in accordance with [15], is

$E_a = m_a \sinh(\lambda/2)/\cosh(\lambda^2/4)$ (14)

Herein $m_a$ is the resource restoration intensity from (9), and $\lambda$ is the parameter of the transferred ultrasound energy.

## 5. Success of Chemotherapy

Success of the chemotherapy in soft matter is determined by the transferred chemical energy in the soft matter and by the entropy in the soft matter. Herein the transferred chemical energy presents the toxicity of chemotherapy, at given its efficacy, and the entropy presents the toxicity of chemotherapy.

The probability $p_{te}$ for transferring of the chemical energy $E_a$ from (14) is defined as the probability with Wigner distribution from [16]. Then this probability is

$p_{te} = (m_n/(2\pi E_a^{1/2}))((E_+ - E_a)(E_a - E_-))^{1/2}$, (15)

$E_- = (2/(100k_s))(m_n + 1) - (4/(100k_s))m_n^{1/2}$, $E_+ = (2/(100k_s))(m_n + 1) + (4/(100k_s))m_n^{1/2}$

Probability $p_{te}$ is the conditional probability of chemotherapy toxicity at given its efficiency.

The entropy of the soft matter h is determined as non-extensive entropy from [17]. Then this entropy is

$h = (3.45/(\pi^{1/2}(100k_s)^{1/2}))G_q((1 + (1/q)(k^*/k_s))^{(-q)} + (1 - (1/q)(k^*/k_s))^q)$, (16)

$G_q = 1/(q^{1/2}\Gamma(q - 1/2)/\Gamma(q) + q^{1/2}\Gamma(q + 1)/\Gamma(q + 3/2))$, $q = m_n$

Herein $\Gamma(.)$ is the gamma function, and $k^*$ is the tumor development intensity.

Prior probability $p_t$ of the chemotherapy toxicity is obtained according to [18]

$\log(p_t) = \pi^2/(6h)$ (17)

Conditional probability for the chemotherapy success in soft matter $p_s$, determined according to Bayes' formula, for prior probability of chemotherapy efficacy $p_e=0.899$, is

$$p_s = 0.899 p_{te}/p_t \qquad (18)$$

6. **Testing of the Success**

The success of the chemotherapy in soft matter is tested for 32 patients with breast cancer. The test shows that the chemotherapy success, referred to the restoration time at chemotherapy, presents distant metastasis-free, given in months, survival from [5]. Here it is assumed that survival is the time scaled success of chemotherapy. Group of patients, with a low-intermediate risk according to 'MKS (mitotic kinome metagene score)', consists of all the patients with a metric tumor marker $m_n=4.1$. Group of patients with a low-intermediate risk according to 'MKS' and a high risk according to '70-gene MammaPrint', includes all patients with a metric tumor marker $m_n=7.3$. Group of patients with a high risk according to 'MKS' and a high risk according to '70-gene MammaPrint' consists only of patients with a metric tumor marker $m_n=2.2$ and $m_n=2.9$.

7. **Conclusion**

The success of chemotherapy in soft matter as a survival is studied in this paper. This success is obtained as follows:

1/ the analogous force of tumor stretching in soft matter is found, through the frequency distribution of the tumor;

2/ ultrasonography is done for this tumor through the change of internal energy of the soft matter;

3/ soft matter restoration with such a tumor is found through the transferred chemical energy at chemotherapy;

4/ Bayes estimate of the probability for chemotherapy success is derived from the transferred chemical energy and from the soft matter entropy;

5/ survival probability is juxtaposed to the time scaling probability of success of chemotherapy.

**References**

1. P G de Gennes, Soft matter (Nobel lecture). Angewandte Chemie International Edition in English, 842-845 (1992)
2. P Coleman, Frontiers at your fingertips. Nature **446**, 379-379 (2007)
3. S Sanga, JP Sinek, HB Frieboes, M Ferrari, JP Fruehauf, V Cristini, Mathematical modeling of cancer progression and response to chemotherapy. Expert Rev. Anticancer Ther. **6**, 1361-1376 (2006)
4. PCW Davies, L Demetrius, JA Tuszynski, Cancer as a dynamical phase transition. Theor. Biol. Med. Model. **8**, 30 (2011)
5. G Bianchini, T Iwamoto, Y Qi, C Coutant, CY Shiang, B Wang, L Santapria, V Valero, GN Hortobagyi, F Symmans, L Gianni, L Pusztai, Prognostic and therapeutic implications of distinct kinase expression patterns in different subtypes of breast cancer. Cancer Res. **70**, 8852-8862 (2010)
6. B Weigelt, J Peterse, LJ van `t Veer, Breast cancer metastasis: Markers and models. Nat. Rev. Cancer **5**, 591-602 (2005)
7. Y Tian, De Sitter thermodynamics from diamond's temperature. J. High Energy Phys. **2005**, 045 (2005)
8. J Cardy, Entanglement entropy in extended quantum systems. arXiv preprint arXiv: cond-mat.stat-mech/0708.2978 (2007)
9. E Canessa, Theory of analogous force on number sets. Phys. A **328**, 44-52 (2003)
10. G Thomas, RS Johal, Expected behavior of quantum thermodynamic machines with prior information. Phys. Rev. E **85**, 041146 (2012)
11. AG Shannon, Generalized Fibonacci matrices in medicine. Notes Number Theory Discrete Math. **15**, 12-21 (2009)
12. SS Masood, Quantum electrodynamics of nanosystems. arXiv preprint arXiv:1204.3770 (2012)
13. MI Hassan, MA Khan, Mathematical modeling as a tool for improving the action of drugs. J. Basic

Appl. Sci. **4,** 81-88 (2008)
14. PP Orth, P Chandra, P Coleman, J Schmalian, Emergent critical phase and Ricci flow in a 2D frustrated Heisenberg model. Phys. Rev. Lett. **109**, 237205 (2012)
15. P Bhattacharya, S Guha, Particle confinement and perturbed dynamical system in warped product spacetime. arXiv preprint arXiv:1202.2328 (2012)
16. M Masuku, JP Rodrigues, How universal is the Wigner distribution? arXiv preprint arXiv: 1107.3681 (2011)
17. MP Leubner, Consequences of entropy bifurcation in non-Maxwellian astrophysical environments. Nonlin. Processes Geophys. **15**, 531-540 (2008)
18. C Carminati, G Tiozzo, Tuning and plateaux for the entropy of α-continued fractions. Nonlinearity **26**, 1049 (2013)